# Complex conductivity of monolayer graphene and Zitterbewegung

*N.E. Firsova, S.A.Ktitorov*

*Ioffe Institute, St. Petersburg, Russia*

A recently derived formula for complex conductivity of monolayer graphene is analyzed. We show that the real and imaginary parts in this formula obey the Kramers-Kronig (KK) dispersion relations which are a good test for validity of the formula for complex conductivity of monolayer graphene. We consider also an additional test for this formula, sensitive to the integral characteristic of the conductance such as the famous f-sum rule. We write it in the two – dimensional form and show that it fulfils identically if we admit the cyclotron mass as an effective one and take the principal value of the integral. We find a deep relation between the graphene complex optical conductivity singularities and electrons Zitterbewegung (ZB) in graphene with frequency $\omega_{ZB}$. Namely, the value of ZB frequency is related with the recently found magnitudes of the inductance $L$ and capacitance $C$ by Thomson's formula i.e. $\omega_{ZB} = 1/\sqrt{LC}$.

## 1. Introduction

The graphene membrane irradiated by the weak time-periodic electric field $E_0 e^{-i\omega t}$ in the terahertz range was considered in [1] to derive the formula for the complex conductance of monolayer graphene square at low temperature $k_B T \ll E_F$

$$Y(\omega) = \frac{e^2}{\hbar}\left\{\frac{1}{4}\Pi\left(\frac{\hbar\omega}{2E_F}\right) - i\frac{1}{\pi}\left[\frac{E_F}{\hbar\omega} + \frac{1}{4}\log\left|\frac{\hbar\omega - 2E_F}{\hbar\omega + 2E_F}\right|\right]\right\} \qquad (1)$$

where $E_F$ is the Fermi energy, $\omega$ is the external field frequency, the symbol $\Pi(\omega)$ is determined as follows:

$$\Pi(x) = \begin{cases} 0 & \text{if } |x| > 1/2 \\ 1/2 & \text{if } |x| = 1/2 \\ 1 & \text{if } |x| < 1/2 \end{cases}. \qquad (2)$$

The method of the derivation of the formula (1) was based on the study of the time-dependent density matrix. The exact solution of the von Neumann equation for density matrix was found within the linear approximation in an external field. The induced current was calculated and then the formula for quantum complex conductivity as a function of the external field frequency, Fermi energy and temperature was derived.

In [2] this formula was obtained with zero real part which corresponds to zero minimal conductivity. However, we see that the minimal conductivity is not equal to zero.

Using (1), there were obtained in [1] also the formulae for capacitance $C$ and inductance $L$ and it was shown there that the graphene membrane was a kind of the resonant circuit near the Dirac point. Namely, the imaginary part of admittance in (1) for $\omega/(2k_F v_F) \ll 1$ has the following form:

$$Y(\omega) = (i\omega L)^{-1} + i\omega C \tag{3}$$

$$L = \frac{\pi \hbar^2}{e^2 E_F} = \frac{\hbar}{e^2} \frac{\pi}{k_F v_F}, \tag{4}$$

$$C = \frac{1}{4\pi} \frac{e^2}{E_F} = \frac{1}{4\pi} \frac{e^2}{\hbar} \frac{1}{k_F v_F}, \tag{5}$$

where $k_F$ and $v_F$ are respectively the Fermi wave number and velocity. So it follows from (1) that near the Dirac point graphene membrane is a kind of parallel resonance circuit. The inductance $L$ and capacitance $C$ are described by formulae (4), (5).

Note that in [3] the suggestion was made that electrons in graphene must exhibit a nonzero mass when collectively excited. Using this notion the inertial acceleration of the electron collective mass and phase delay of the resulting current were considered. On the basis of this model the so-called kinetic inductance representing the reluctance of the collective mass to acceleration was introduced, calculated and measured. The obtained expression for inductance coincides with the formula (4) for the unity width $w$.

Analyzing the formula for inductivity we see that $L \sim n^{-1/2}$ that means that the less is the carrier density the more is the inductivity. Just such dependence was observed in the experiment [3].

The formula for quantum capacitance based on the two-dimensional free electron gas model was suggested in [4]. Notice that in [5] the graphene membrane quantum capacitance was firstly measured. The expression for capacitance obtained in [4] coincides with (5) being applied to the unity square up to the numerical factor $8\pi$. We see that the graphene membrane eigenfrequency (conductivity singularity) reads:

$$\omega_{res} = 2E_F / \hbar = 2k_F v_F, \qquad \lim_{\omega \to \omega_{res}} \sigma(\omega) = \infty$$

(6)

We see also from the formulae (4), (5) that for the found resonant frequency we have the usual equation

$$\omega_{res} = 1/\sqrt{LC},$$

(7)

which means that near the Dirac point the graphene membrane is a resonant circuit with the eigenfrequency $\omega_{res}$. Notice that the conductivity singularity in the point $\omega_{res}$ is logarithmic instead of the pole-type we used to see in 3D. We think that it is the 2D-effect.

Note that the formulae for quantum inductance and capacitance from [3-5] actually were obtained on the basis of different models. So if we used these formulae from [3-5] to find the eigenfrequency according to Thompson's formula (7), we would have got the value where the conductivity (1) does not have singularity. It means that the formulae (4), (5) for the inductance and capacitance are consistent since they give a correct value of the resonance frequency calculated with a use of the Thomson formula, which corresponds to the singularity of the conductance.

Note also that if the condition $\omega/(2k_F v_F) \ll 1$ is not fulfilled (which we have for low carrier density), then the quantum capacitance would depend on frequency as follows (see [1])

$$C(\omega) = \frac{e^2}{4\pi\hbar\omega}\log\left|1 + 2\frac{\hbar\omega}{2E_F - \hbar\omega}\right| > 0$$

(8)

while $L$ does not depend on $\omega$.

## 2. Quantum admittance of the graphene

### A. Kramers-Kronig dispersion relations.

We show that the found in [1] real and imaginary parts of complex conductivity obey the Kramers-Kronig (KK) dispersion relations:

$$Y(\omega) = Y'(\omega) + iY''(\omega)$$
$$Y'(\omega) = -\frac{1}{\pi} P \int_{-\infty}^{\infty} dx \frac{Y''(x)}{x - \omega}, \quad Y''(\omega) = \frac{1}{\pi} P \int_{-\infty}^{\infty} dx \frac{Y'(x)}{x - \omega}.$$

(9)

In order to make the formula (1) to be consistent with (9), one has to add the Drude peak to the real part of conductance:

$$Y'(\omega) = \frac{e^2}{4\hbar} \Pi\left(\frac{\hbar\omega}{2E_F}\right) + D\delta\left(\frac{\hbar\omega}{2E_F}\right),$$

(10)

where the Drude weight is determined as follows

$$D = \frac{e^2}{\hbar} E_F / \hbar.$$

(11)

The Kramers – Kronig relations are good tests for validity of the conductance formula and following from it resonance frequency, which is determined by the product *LC*. At the same time *L* and *C* are checked up to an arbitrary numerical factor. This problem will be considered in the next subsection.

### B. Sum rule.

Now we will consider additional verification of these unchecked numerical factors using a formula sensitive to the integral characteristic of the conductance such as the famous Thomas-Reiche-Kuhn f – sum rule valid for three-dimensional metals [6]

$$2\int_0^\infty d\omega \lim_{q\to\infty} \sigma(\mathbf{q},\omega) = \frac{\pi e^2 n}{m_0}, \qquad (12)$$

where $\sigma(\mathbf{q},\omega)$ and $n$ are respectively the 3d conductivity and the three-dimensional electron density, $m_0$ is the bare electron mass in the free space.

This sum rule is not directly applicable to two dimensional zero-gap systems like graphene [7, 8]. Known relevant formulae derived in [7, 8] require passing from the Dirac model to more sophisticated one: a transfer to the tight-binding approximation or introduction of the energy cut-off. The following formula was derived in [7]:

$$\int_0^{\Lambda_E} d\omega \cdot \omega \operatorname{Im} \chi(\mathbf{q},\omega) = -q^2 \Lambda_E / 4, \qquad (13)$$

where $\chi(\mathbf{q},\omega)$ is the density – density correlator, $\Lambda_E$ is the frequency cut-off. This formula is of use for only qualitative check of the response function since the cut-off is not precisely determined value. Unfortunately, we have not yet succeeded in consisted derivation of a relevant relation for graphene. However, we advance here a hypothesis that a role of the electronic mass in the sum rule for graphene can play the cyclotron mass $m_c = E_F / v_F^2 = \frac{\hbar k_F}{v_F}$. Useful discussion of this problem can be found in [2].

Introducing the sheet conductance $G_s(\mathbf{q},\omega)$ and the two-dimensional electron density $N/S$, we can rewrite equation (12) in the form

$$2\int_0^\infty d\omega \operatorname{Re} \lim_{\mathbf{q}\to 0} G_S(\mathbf{q},\omega) = \frac{\pi e^2 N/S}{m_c}. \qquad (14)$$

The mean two-dimensional electron density can be expressed in terms of the Fermi momentum by the relation:

$$N/S = \nu \frac{2\pi}{(2\pi\hbar)^2} \int_0^{P_F} p\,dp = \nu \frac{p_F^2}{\hbar^2 4\pi} = \frac{p_F^2}{\hbar^2 \pi}, \quad (15)$$

where $\nu = 4$ takes account of the spin and valley degeneracy. Substituting (15) and (1) into (14) we see that the equation (13) turns into identity if we admit the cyclotron mass as an effective one and take the principal value of the integral in the left side of formula (14) excluding a contribution of the Drude peak. So the formula for conductance is shown to satisfy the two-dimensional sum rule. The fact that the sum rule is satisfied means the validity of formulae (1, 4, 5).

And vice versa, if we were sure of the conductance formula validity, this fact could verify our sum rule formula.

### C. Electrons Zitterbewegung in graphene.

We find a deep relation between the graphene complex optical conductivity peculiarities and Zitterbewegung (ZB). ZB ("trembling motion" from German) is a fast oscillating motion of elementary particles, in particular electrons that obey the Dirac equation. ZB phenomenon was first predicted by Erwin Schrödinger in 1930 [9] as a result of
his analysis of the wave packet solutions of the Dirac equation for electrons in free space. He concluded that the interference between positive and negative energy states produced what appeared to be a fluctuation (at the speed of light) of the position of an electron around the mean value with the frequency $2mc^2/\hbar$. ZB of a free relativistic particle has never been observed because of the huge value of the trembling frequency.

However, graphene with its Dirac states near the K and K' critical points is an ideal test area to simulate many of the quantum electrodynamics phenomena. In particular, ZB was shown in [10] to exist in graphene with the trembling frequency realistic to be measured.

Also we see that the complex conductivity obtained in [1] has a resonance at the frequency, which coincides with the ZB one $\omega_{ZB}$ calculated in [10] for monolayer graphene.

On the other hand we show that the resonance frequency can be expressed through the introduced in [1] values for the graphene inductance and capacitance. We see also that the value of ZB frequency is related with the found in [1] magnitudes of the inductance and capacitance by Thomson's formula i.e.

$$\omega_{ZB} = \frac{1}{\sqrt{LC}} \quad (16)$$

**Remark.** Note that the presented above formulae are valid in units with the speed of light $c$ equal to unity. Introducing new variable $\tilde{L} = c^2 L$ for the inductance, we obtain the standard relations in the Gauss units:

$$\omega_{ZB} = \frac{c}{\sqrt{\tilde{L}C}}$$

In order to come to SI units one should multiply $\tilde{L}$ and $C$ by the electrodynamic constants:

$$\tilde{L} \to \tilde{L}\mu_0, \quad C \to C\varepsilon_0$$

## 3. Conclusions

We analyzed the formula for the graphene complex electroconductance derived in [1]. We have shown that the real and imaginary parts in this formula obey the Kramers-Kronig (KK) dispersion relations, which is a good test for validity of this formula for complex conductivity of monolayer graphene. We have considered also additional test for this formula, sensitive to the integral characteristic of the conductance such as the famous f-sum rule. We have written it in the two-dimensional form and have demonstrated that it would be fulfilled identically if we admit the cyclotron mass as an effective one and take the principal value of the integral.

We have found a deep relation between the graphene complex optical conductivity singularities and electrons Zitterbewegung (ZB) in graphene with the frequency $\omega_{ZB}$. Namely, as it was shown in [1] the frequency dependence of the admittance near the Dirac point is similar to one for the parallel resonance circuit. This result allowed in [1] to calculate corresponding capacitance and inductance for graphene. The eigenfrequency of the circuit obtained in [1] appeared to be equal to the Zitterbewegung frequency i.e. the value of ZB frequency is related with the found in [1] magnitudes of the inductance $L$ and capacitance $C$ by Thomson's formula.

So we see that it is just ZB which determines conductivity dynamics in the vicinity of the Dirac point. So we can say that the main sense of the formula for complex conductivity in the vicinity of the Dirac point derived in [1] is ZB. The suggested sum rule allows us to verify the formula for conductance (1).

The obtained results are important for nanoelectronics including plasmonics, radiation sensitive devices, terahertz radiation amplification and generation. This study is also useful for understanding of ultra high energy processes in quantum electrodynamics.

## References

1. Firsova, N.E. *Photonics and Nanostructures – Fundamentals and Applications*. **2017**, *26*, 8-14.
2. Gusynin, V.P.; Sharapov, S.G.; Carbotte. *Phys. Rev*. B, **2007**, *75*, 165407.
3. Yoon Hosang, Forsythe Carlos et al. *Nature nanotechnology*. **2014**, *9*, 594-599.
4. Fang, T., Konar, A., Xing, H. L, Jena, D. *Appl. Phys. Lett*. **2007**, *91*, 092109.
5. J. Xia Jilin, F. Chen, et al, *Nature nanotechnology*. **2009**, *4,* 505.
6. Pines D., *Elementary excitations in solids*. W.A. Benjamin Incorporated: New York - Amsterdam, 1963.
7. Sabio, J.; Nilsson, 1 J.; Castro Neto, A. H. *Phys. Rev*. B, **2008**, *78*, 075410 .
8. Throckmorton, R. E.; Das Sarma, S., *Phys. Rev*. B, **2018**, *98*, 155112.
9. Schrödinger, E., *Berliner Ber*. **1930**, 418.
10. M. I. Katsnelson, *The European Physical Journal.* B  **2006**, *51*, 157.